\documentclass[aps,prd,twocolumn,10pt
,
tightenlines,superscriptaddress,
nofootinbib,showpacs]{revtex4-1}
\usepackage{amssymb,latexsym}
\usepackage{amsmath,amsbsy,bbm}
\usepackage{epsfig,bm}
\usepackage{graphicx,comment}
\newcommand\beq{\begin{eqnarray}}
\newcommand\eeq{\end{eqnarray}}
\newcommand\nn{\nonumber}

\begin{document}
\def\vp{{\vec{p}}}
\def\vx{{\vec{x}}}
\def\vD{{\vec{D}}}
\def\vB{{\vec{B}}}
\def\vE{{\vec{E}}}
\def\vcE{{\vec{\mathcal{E}}}}
\def\vA{{\vec{A}}}
\def\vDel{{\vec{\nabla}}}
\def\vzero{{\vec{0}}}

\def\a{{\alpha}}
\def\b{{\beta}}
\def\d{{\delta}}
\def\D{{\Delta}}
\def\X{{\Xi}}
\def\e{{\varepsilon}}
\def\g{{\gamma}}
\def\G{{\Gamma}}
\def\k{{\kappa}}
\def\l{{\lambda}}
\def\L{{\Lambda}}
\def\m{{\mu}}
\def\n{{\nu}}
\def\o{{\omega}}
\def\O{{\Omega}}
\def\S{{\Sigma}}
\def\s{{\sigma}}
\def\th{{\theta}}

\def\ol#1{{\overline{#1}}}

\def\Dslash{D\hskip-0.65em /}
\def\diag{\text{diag}}

\def\cE{{\mathcal E}}
\def\cF{{\mathcal F}}
\def\cG{{\mathcal G}}
\def\cS{{\mathcal S}}
\def\cC{{\mathcal C}}
\def\cB{{\mathcal B}}
\def\cT{{\mathcal T}}
\def\cQ{{\mathcal Q}}
\def\cL{{\mathcal L}}
\def\cO{{\mathcal O}}
\def\cA{{\mathcal A}}
\def\cQ{{\mathcal Q}}
\def\cR{{\mathcal R}}
\def\cs{{\mathfrak s}}
\def\cH{{\mathcal H}}
\def\cW{{\mathcal W}}
\def\cM{{\mathcal M}}
\def\cD{{\mathcal D}}
\def\cN{{\mathcal N}}
\def\cP{{\mathcal P}}
\def\cK{{\mathcal K}}

\def\eqref#1{{(\ref{#1})}}

\title{Background Electromagnetic Fields and NRQED Matching: Scalar Case}

\author{Jong-Wan~Lee}
\email[]{jwlee2@ccny.cuny.edu}
\affiliation{
Department of Physics,
        The City College of New York,  
        New York, NY 10031, USA}
\author{Brian~C.~Tiburzi}
\email[]{btiburzi@ccny.cuny.edu}
\affiliation{
Department of Physics,
        The City College of New York,  
        New York, NY 10031, USA}
\affiliation{
Graduate School and University Center,
        The City University of New York,
        New York, NY 10016, USA}
\affiliation{
RIKEN BNL Research Center, 
        Brookhaven National Laboratory, 
        Upton, NY 11973, USA}
\date{\today}

\pacs{12.39.Hg, 13.40.Gp, 13.60.Fz, 14.20.Dh}

\begin{abstract}
The low-energy structure of hadrons can be described systematically using effective field theory, 
and the parameters of the effective theory can be determined from lattice QCD computations. 
Recent work, 
however, 
points to inconsistencies between the background field method in lattice QCD and effective field theory matching conditions. 
We show that the background field problem necessitates inclusion of operators related by equations of motion. 
In the presence of time-dependent electromagnetic fields, 
for example, 
such operators modify Green's functions, 
thereby complicating the isolation of hadronic parameters which enter on-shell scattering amplitudes. 
The particularly simple case of a scalar hadron coupled to uniform electromagnetic fields is investigated in detail.
At the level of the relativistic effective theory, 
operators related by equations of motion are demonstrated to be innocuous. 
The same result does not hold in the non-relativistic effective theory, 
and inconsistencies in matching are resolved by carefully treating operators related by equations of motion.  
As uniform external fields potentially allow for surface terms, 
the problem is additionally analyzed on a torus where such terms are absent. 
Finite-size corrections are derived for charged scalar correlation functions in uniform electric fields as a useful byproduct. 
\end{abstract}
\maketitle

\section{Introduction} %

The electromagnetic structure of hadrons can be determined directly from photon-hadron scattering cross sections,
but also affects low-energy quantities measured in high-precision experiments. 
A prominent example that has sparked considerable recent interest is the proton charge radius, 
which is a quantity appearing in the low momentum-transfer expansion of the proton 
electric form factor. 
While the proton charge radius can be extracted from electron-proton scattering cross sections, 
this quantity additionally gives rise to the leading finite-size effect in the spectrum of hydrogenic atoms. 
From high-precision measurements of the muonic hydrogen spectrum%
~\cite{Pohl:2010zza,Antognini:1900ns}, 
the extracted proton radius is discrepant with that from scattering data at the 
$7 \sigma$
level, 
see~\cite{Pohl:2013yb} for a comprehensive review.%
\footnote{ 
Another example, 
which is directly related to the present work, 
concerns the electromagnetic structure of the pion. 
Charged pion electromagnetic polarizabilities determined from chiral perturbation theory%
~\cite{Holstein:1990qy}
are discrepant with scattering experiments by a factor of two~\cite{Ahrens:2004mg}, 
which corresponds to 
$\sim 2.5 \sigma$.
These low-energy quantities also appear in high-precision physics, 
namely as hadronic corrections to the anomalous magnetic moment of the muon%
~\cite{Engel:2012xb,Engel:2013kda}.
}

The systematic and unified treatment of low-energy hadron structure is afforded by effective field theory techniques. 
The description of proton-size effects in non-relativistic quantum electrodynamics (NRQED), 
for example,  
has been given in%
~\cite{Hill:2011wy}.
In the effective hadronic theory, 
low-energy interactions are systematically written down with parameters that encompass
hadronic structure.
Effective field theory matching allows one to relate the universal low-energy parameters to physical observables. 
In this way, 
one sees that the same parameters which enter the description of scattering cross sections, 
also enter high-precision low-energy quantities, 
such as proton-size corrections to the muonic hydrogen spectrum. 
In principle, 
the parameters entering the effective hadronic theory can be computed using lattice QCD, 
and steps in this direction have been made. 
The present work concerns the extension of effective field theory matching to the case of background fields. 
To be clear, 
we find no problems with effective field theory matching of 
$S$-matrix elements, 
however, 
the extension of effective field theory matching to theories in background fields involves a subtlety.

Background field calculations in lattice QCD represent a fruitful method to determine hadronic properties, 
see~\cite{Martinelli:1982cb,Bernard:1982yu,Fiebig:1988en,Lee:2005ds,Christensen:2004ca,Lee:2005dq,Shintani:2006xr,Shintani:2008nt,Engelhardt:2007ub,Detmold:2009dx,Alexandru:2009id,Detmold:2010ts,Alexandru:2010dx,Freeman:2012cy,Primer:2013pva,Lujan:2013qua}. 
In particular, 
electromagnetic polarizabilities can be accessed using the background field method on the lattice, 
while photon-hadron scattering computations are beyond the current and foreseeable reach of lattice QCD. 
There is a theoretical need to understand the relation between parameters extracted from background field lattice calculations and those reported by the Particle Data Group. 
In this respect, 
various groups calculating the neutron electric polarizability, 
for example,  
are not determining the same quantity. 
Before this issue can be addressed, 
we must first understand how to match effective theories in external fields.

In the present work, 
we expose a subtlety in matching hadronic effective field theories in electromagnetic fields. 
To highlight this subtlety, 
let us point to an inconsistency that results from incorrect matching conditions in external fields. 
Applying the NRQED method of~\cite{Hill:2012rh}
to determine the initial energy shift of a charged scalar in a uniform electric field, 
we obtain the result
(to be discussed in Sec.~\ref{s:EHQET} below)
\begin{equation}
\Delta E
= 
- \frac{1}{2} 
\left[
4 \pi \alpha_E 
- 
\frac{Z}{3 M}
< r^2> 
\right]
\vE^2
\label{eq:one}
,\end{equation}
where 
$M$, 
$\alpha_E$, 
and 
$<r^2>$
are the scalar's mass, 
electric polarizability, 
and charge radius, 
respectively. 
Appearance of the charge radius in the energy shift is rather surprising, 
because such virtual photon contributions should be absent on strictly physical grounds. 
This result is to be contrasted with the initial energy shift in the relativistic case~\cite{Detmold:2009dx}
\begin{equation}
\Delta E 
= 
- \frac{1}{2} 4 \pi \alpha_E \vE^2
\label{eq:twoo}
,\end{equation}
which turns out to be the correct result.%
\footnote{
Here is where the exact definition of the initial energy becomes important. 
Technically we must determine the non-relativistic expansion of the relativistic correlator 
to compare the initial energy with that appearing in Eq.~\eqref{eq:one}. 
Performing this expansion 
(also to be discussed in Sec.~\ref{s:EHQET} below), 
we obtain the non-relativistic initial energy shift
$\Delta E = - \frac{1}{2} \left[ 4 \pi \alpha_E + \frac{Z^2}{2 M^3} \right] \vE^2$. 
Notice that the additional term in the energy shift is already necessarily contained in the relativistic correlator employed in the analysis of%
~\cite{Detmold:2009dx}. 
 } 
Resolution of this inconsistency is one of the goals of this work. 
We find that resolution is possible by extending matching to the Green's functions, 
which requires retaining effective field theory operators related by equations of motion. 
As a result,  
the matching of $S$-matrix elements determined in~\cite{Hill:2012rh} is completely unaffected, 
but can be modified in external fields. 
The modification accounts for the difference between 
Eqs.~\eqref{eq:one}
and \eqref{eq:twoo}.

Throughout we consider the dynamics of a composite scalar coupled to electromagnetic fields, 
and our presentation is organized as follows. 
We begin in Sec.~\ref{s:demo} with a demonstration that operators related by equations of motion can modify Green's functions. 
For time-dependent electromagnetic fields, 
we provide an illustrative example that points to an obstruction in the extraction of on-shell properties using background field correlators in lattice QCD. 
Despite this general obstruction, 
we show in Sec.~\ref{s:scalar}
that the particular case of a charged, relativistic, scalar particle coupled to uniform electromagnetic fields happens to be immune to such difficulties. 
To facilitate matching with the non-relativistic theory,  
we additionally compute one- and two-photon processes to relate the parameters of the relativistic theory to observables, 
and obtain the correlation functions of charged relativistic scalars in uniform electric and magnetic fields. 
Next in Sec.~\ref{s:NRscalar}, 
we write down the non-relativistic theory of a composite scalar using HQET power counting.%
\footnote{
Here we employ the acronym HQET for any effective theory that utilizes an expansion in 
inverse powers of a particle's mass.
This power counting is, 
of course, 
shared by its namesake, 
heavy quark effective theory. 
} 
 We argue for the inclusion of an additional operator which ordinarily would be eliminated by use of the equations of motion. 
The operator is shown to modify charged scalar Green's functions in uniform electric fields, 
and is required so that Green's functions match between relativistic and non-relativistic theories. 
To compute the non-relativistic propagator,
we employ both HQET and NRQED power counting, 
and obtain results consistent with the non-relativistic reduction of the relativistic theory only when 
operators related by equations of motion are retained in the non-relativistic theory. 
As a final check, 
we expand the relativistic theory in powers of the scalar's mass for a brute-force determination of the matching coefficients at the level of the action. 
Certain technical details are relegated to Appendices. 
The problem of a charged scalar hadron in an electric field is formulated on a Euclidean torus in 
Appendix~\ref{s:B},
and precludes the possibility of surface terms. 
The NRQED expansion of the relativistic charged scalar propagator is determined in 
Appendix~\ref{s:A}. 
In the last section of the main text, 
Sec.~\ref{s:S}, 
we summarize our findings.

\section{Operators Related by Equations of Motion} \label{s:demo} %

Ordinarily operators related by equations of motion are redundant, 
and can be dropped from an effective field theory. 
This has the desirable consequence of reducing the number of low-energy parameters, 
which is essential in economically parameterizing the model-independent physics relevant at a given energy scale. 
In external fields, 
however, 
the issue becomes subtle, 
and our goal is to expose the subtlety first in the context of a simplified example.

Consider the following toy-model Lagrange density for a  charged composite scalar
\begin{eqnarray}
\cL 
&=&
D_\mu 
\Phi^\dagger 
D^\mu 
\Phi
-
M^2
\Phi^\dagger
\Phi
+
\frac{C}{2 M^4}
\Phi^\dagger \Phi 
\,
\partial^2 F^2
\notag \\
&&
+
\frac{\mathfrak{C}}{M^4}
F^2
\left(
D_\mu 
\Phi^\dagger 
D^\mu 
\Phi
- 
M^2 
\Phi^\dagger \Phi
\right)
\label{eq:toy}
,\end{eqnarray} 
where 
$F_{\mu \nu}$
is the electromagnetic field-strength tensor, 
with
$C$
and 
$\mathfrak{C}$
as the dimensionless low-energy constants of this model. 
Non-minimal photon couplings are allowed because we assume 
$\Phi$
is a composite particle with charged constituents. 
The electromagnetic gauge covariant derivatives have the action
\beq
D_\mu \Phi 
&=& 
\partial_\mu \Phi + i Z A_\mu \Phi,
\notag \\ 
D_\mu \Phi^\dagger 
&=& 
\partial_\mu \Phi^\dagger - i Z A_\mu \Phi^\dagger
.\eeq
No power counting has been utilized in writing
Eq.~\eqref{eq:toy}; 
in this section, 
we merely select operators to illustrate our point. 
It will prove useful to treat the electromagnetic coupling as small. 
To this end, 
we consider the parameters
$C$ 
and 
$\mathfrak{C}$
to be proportional to the square of the electric charge, 
$\alpha = \frac{e^2}{4 \pi}$, 
and we will drop terms of 
$\cO(\alpha^2)$
in what follows.

For processes with only on-shell
$\Phi$
states, 
it turns out that observables, 
such as the amplitude for the Compton scattering process 
$\gamma + \Phi \to \gamma + \Phi$, 
depend only on a particular linear combination of low-energy parameters,
$C + \mathfrak{C}$. 
For virtual 
$\Phi$
states, 
the diagrammatic analysis is more involved; 
but, 
off-shell contributions from the equation-of-motion operator generally can be removed in the renormalization of the theory. 
Because the diagrammatic approach is cumbersome, 
we handle the removal of redundant operators by employing field redefinitions, 
see~\cite{Arzt:1993gz}
and references therein. 
For the toy model, 
we invoke the field redefinition
\begin{equation}
\Phi 
=
\left( 
1 
-
\frac{\mathfrak{C}}{2 M^4} F^2
\right)
\Phi'
\label{eq:FRD}
,\end{equation}
which corresponds to dressing the scalar field with photons. 
After field redefinition, 
the theory is described by the Lagrange density
\begin{eqnarray}
\cL'
&=&
D_\mu \Phi' D^\mu \Phi' 
-
M^2 
\Phi'^\dagger
\Phi'
+
\frac{C'}{2 M^4}
\Phi'^\dagger \Phi' \partial^2 F^2
+ \cO(\alpha^2),
\notag \\
\label{eq:reduced}
\end{eqnarray}
with 
$C' = C + \mathfrak{C}$. 
The operator related by equations of motion has now been removed. 
The coefficient 
$C'$
can be chosen so that Eq.~\eqref{eq:reduced} reproduces 
$S$-matrix elements for processes involving 
the composite scalar and photons. 
This procedure exposes that on-shell processes depend only on  
$C'$. 
In Eq.~\eqref{eq:toy}, 
additional dependence on the parameter 
$\mathfrak{C}$
that can arise from virtual 
$\Phi$ 
contributions in loop diagrams
must be cancelled by the counter-terms necessary to renormalize the theory. 
In this way, 
the theories described by
Eqs.~\eqref{eq:toy} 
and
\eqref{eq:reduced}
are equivalent.

Now consider the toy-model Lagrange density for the case where 
$F_{\mu \nu}$
is a time-dependent external field.
The explicit time-dependence introduced eliminates the possibility of an on-shell condition. 
As a result, 
one cannot appeal to a renormalization prescription to fix the behavior of the 
two-point function at the single-particle pole, 
because there are no such poles. 
Consequently the parameters 
$C$ 
and 
$\mathfrak{C}$ 
can be resolved at the level of the Green's function.%
\footnote{
Notice that the charged particle Green's function is gauge dependent. 
Implicitly included in the choice of external field is the gauge, 
which is then fixed. 
We will derive results below for particularly simple gauge choices; 
results in other gauges can similarly be derived. 
While appending an electromagnetic gauge link between operators in the two-point function will lead one to gauge invariant Green's functions, 
these Green's functions will then depend on the path chosen to link the operators. 
Path dependence arises because flux threads loops transverse to the electromagnetic fields. 
} 
Suppose we start with the reduced theory described by 
Eq.~\eqref{eq:reduced}.  
The propagator for 
$\Phi'$
we write as
$\cG'(x,y)$, 
with
\begin{equation} \label{eq:propsimple}
\cG'(x,y) 
=
\langle 0 | T\left\{ \Phi' (x) \Phi'^\dagger(y) \right\} | 0 \rangle 
.\end{equation} 
Starting with the theory in Eq.~\eqref{eq:toy}, 
on the other hand, 
the propagator for 
$\Phi$
we write as 
$\cG(x,y)$.
This propagator can be deduced simply%
\footnote{
One can also compute the 
$\Phi$
propagator directly by treating the operator with coefficient
$\mathfrak{C}$
in 
Eq.~\eqref{eq:toy}
as a perturbation. 
In this approach, 
one works in coordinate space, 
and utilizes the Green's function identity 
$( D^2_y + M^2) \cG(x,y) = i \delta(x-y)$. 
The resulting propagator is the same as 
Eq.~\eqref{eq:propmess}, 
and eliminates possible additional factors that could appear from carrying out the field redefinition carefully at the level of the functional integral. 
} 
by utilizing the field redefinition in 
Eq.~\eqref{eq:FRD}
\begin{eqnarray}
\cG(x,y)
&=& 
\langle 0 | T\left\{ \Phi(x) \Phi^\dagger(y) \right\} | 0 \rangle
\notag \\
&=&
\left[
1 
- 
\frac{\mathfrak{C}}{2 M^4} 
[
F^2(x) 
+ 
F^2(y)
]
\right]
\cG'(x,y),
\label{eq:propmess}
\end{eqnarray}
where we have dropped contributions that are of order 
$\alpha^2$. 
Notice this correlator necessarily has different time dependence.%
\footnote{
It is useful to imagine the case of a time-independent magnetic field, 
for which one has an on-shell condition. 
In this case, 
the operator with coefficient 
$\mathfrak{C}$
does not modify the spectrum of the theory as can be shown by taking the temporal Fourier transform of 
Eq.~\eqref{eq:propmess}, 
\begin{equation}
\cG(\vec{x}, \vec{y} \, | E) 
= 
\left[
1 
- 
\frac{\mathfrak{C}}{2 M^4} 
[
F^2(\vec{x} \,) 
+ 
F^2(\vec{y} \,)
]
\right]
\cG'(\vec{x},\vec{y} \, | E )
\notag
.\end{equation}
The residues at each energy pole, 
however, 
are different between the reduced and unreduced theories.  
This difference reflects perturbative corrections to the coordinate wavefunctions of energy eigenstates. 
Without the explicit coordinate dependence introduced by the magnetic field in this case, 
one could impose the standard wavefunction renormalization condition which would lead to on-shell Green's functions that match, both poles and residues.
}

The difference between propagators has an important consequence for the background field method in lattice QCD computations. 
After computing correlation functions of the scalar particle on the lattice, 
we must match the behavior of the lattice-determined correlator with the prediction from an effective hadronic theory. 
Without an on-shell condition, 
this step is essential because the method hinges on the effective theory being able to reproduce the time dependence of the lattice correlator data. 
We cannot simply assume that the external field propagator will be given by
Eq.~\eqref{eq:propsimple}. 
The corresponding effective theory has been reduced by a field redefinition. 
The most general effective theory should start with all possible operators, 
including operators that one would normally remove via field redefinitions. 
Such a general theory will have different Green's functions than its corresponding reduced theory; 
and, 
in this way, 
two such theories are hence no longer equivalent.

From the time dependence of the Green's function, 
parameters associated with on-shell particles can be extracted. 
In our toy-model example, 
one has access to both parameters
$C'$
and 
$\mathfrak{C}$
from the propagator, 
Eq.~\eqref{eq:propmess}. 
In the corresponding effective theory, 
Eq.~\eqref{eq:toy}, 
only the parameter 
$C'$
contributes to on-shell properties of
$\Phi$. 
Thus when considering time-dependent external field correlation functions, 
we must retain operators ordinarily removed by the equations of motion. 
Such operators affect the time dependence of Green's functions, 
and their coefficients are generally not related to physical properties of the particle. 
In light of this observation, 
we consider effective theories of relativistic and non-relativistic charged scalars 
in uniform electric and magnetic fields, 
and address possible contributions from operators related by the equations of motion.

\section{Relativistic Scalar QED} \label{s:scalar} %

Moving on from the toy-model example, 
we consider the fully relativistic action for a charged composite scalar
$\Phi$
interacting with electromagnetic fields. 
This provides an effective hadronic theory that will ultimately be reduced to a non-relativistic theory below, 
and has useful applications to pions and Helium--$4$. 
We detail the case of the scalar propagating in uniform electromagnetic fields;
and, 
from our discussion above, 
we need to address possible contributions arising from operators related by the equations of motion.

\subsection{Operators}
To write down the effective theory for 
$\Phi$, 
we accordingly enforce upon the action the usual 
$C$, 
$P$, 
and 
$T$
invariance in addition to Lorentz and gauge invariance. 
The operators of this theory can be organized in powers of increasing mass dimension, 
however, 
we do not write down all possible terms up to dimension eight. 
Instead, 
we keep only terms which turn out to be relevant in the non-relativistic limit, 
specifically up to 
$\mathcal{O}(M^{-4})$. 
Writing down all of those such terms, 
we have
\begin{eqnarray}
\cL
&=&
D_\mu \Phi^\dagger D^\mu \Phi 
-
M^2 
\Phi^\dagger \Phi
\notag\\
&&
- 
\frac{C_0}{M^2} \,
F^2 \Phi^\dagger \Phi
+
\frac{C_1}{M^2} \,
[\partial_\mu F^{\mu \nu}] J_\nu
\notag\\
&&
+
\frac{C_2}{M^4} \,
T_{\mu \nu} D^\mu \Phi^\dagger D^\nu \Phi
-
\frac{C_3}{M^4} \,
[\partial^2 \partial_\mu F^{\mu \nu}] 
J_\nu
\label{eq:rel}
.\end{eqnarray}
We employ 
$T_{\mu \nu}$
for the electromagnetic stress-energy tensor, 
$T_{\mu \nu} = F_{\rho \{ \mu } F_{  \nu \} } {}^\rho$, 
where the curly braces denote symmetrization and trace subtraction, 
\beq
\cO_{ \{ \mu \nu \} } = \frac{1}{2} \left( \cO_{\mu \nu} + \cO_{\nu \mu} - \frac{1}{2} g_{\mu \nu} \cO_\a {}^\a \right)
.\eeq 
The vector current is given by
\beq
J_\mu = i \left( \Phi^\dagger [D_\mu \Phi] - [D_\mu \Phi^\dagger] \Phi \right)
,\eeq 
where, 
in order to remove ambiguity throughout, 
we have adopted the convention that bracketed derivatives only act inside the square brackets.%
\footnote{ 
To obtain the theory corresponding to a neutral scalar hadron, 
such as the neutral pion, 
one sets 
$Z = 0$, 
imposes 
$\Phi^\dagger = \Phi$, 
and includes the correct normalization factors.
Accordingly the operators with coefficients
$C_1$
and
$C_3$
vanish, 
and the theory only depends on 
$C_0$
and
$C_2$. 
Finally due to isospin breaking, 
the values of coefficients in the neutral hadron theory are generally unrelated to those in the charged hadron theory.  
}

In writing down the effective theory in 
Eq.~\eqref{eq:rel}, 
we assumed the absence of surface terms to eliminate operators. 
At dimension six, 
for example, 
we utilized the identity
\begin{eqnarray}
\partial_\mu 
\left( 
F^{\mu \nu} J_\nu
\right)
&=&
[\partial_\mu F^{\mu \nu}] J_\nu
+ 
2 i
F^{\mu \nu} D_\mu \Phi^\dagger D_\nu \Phi 
\notag \\
&&
+ 
Z 
F^2 \Phi^\dagger \Phi
,\end{eqnarray}
to exclude the operator
$i F^{\mu \nu} D_\mu \Phi^\dagger D_\nu \Phi$.  
For a uniform external field extending over all spacetime, 
it is not immediately obvious that surface terms vanish due to the linearly rising four-vector potential
$A_\mu$
required to obtain such electromagnetic fields. 
For lattice applications, 
we restrict such electromagnetic fields to a Euclidean torus in Appendix~\ref{s:B}. 
In that case, 
there are no surface terms in both relativistic and non-relativistic theories, 
and infinite spacetime results are recovered exponentially fast in the spacetime volume.

The relativistic theory in Eq.~\eqref{eq:rel} has been written,
moreover,  
without any operators related by equations of motion. 
For spacetime varying external fields,
the effect of such operators on Green's functions is generally rather complicated.  
Ultimately we are concerned with determining the behavior of Green's functions in the effective theory to compare with lattice QCD data computed in uniform electric and magnetic fields. 
To this end, 
we derive specific results for the case of uniform electric and magnetic fields. 
Unlike the toy-model example above,
there are valuable simplifications in this case.

For external electromagnetic fields, 
the additional terms required in the effective theory appear in the Lagrange density
\begin{eqnarray}
\Delta \cL
&=&
\frac{\mathfrak{C}_0}{M^2}
\Phi^\dagger (D^2 + M^2)^2 \Phi
+
\frac{\mathfrak{C}_1}{M^4}
\Phi^\dagger (D^2 + M^2)^3 \Phi
\notag \\
&&
+
\frac{\mathfrak{C}_2}{M^4}
F^2
\Phi^\dagger (D^2 + M^2) \Phi
.\label{eq:eomOs}
\end{eqnarray}
These operators have been simplified using integration by parts. 
In particular, 
an integration by parts shows that the last operator 
is equivalent to the equation-of-motion operator appearing in the toy-model Lagrange density, 
Eq.~\eqref{eq:toy}, 
under the assumption that the electromagnetic fields are uniform.
The operators appearing in Eq.~\eqref{eq:eomOs} 
can be removed with a field redefinition having the form
\begin{eqnarray}
\Phi 
&=& 
\Bigg[ 
1 
+ 
\frac{\mathfrak{C}_0}{2 M^2}
( D^2 + M^2)
\notag \\
&&
+
\frac{\mathfrak{C}_1}{2 M^4}
(D^2 + M^2)^2
+ 
\frac{\mathfrak{C}_2 - C_0 \mathfrak{C}_0}{2 M^4}
F^2
\Bigg]
\Phi'
\label{eq:relFRD}
.\end{eqnarray}
In terms of the redefined field, 
we accordingly have
\begin{eqnarray}
\cL + \Delta \cL 
&=&
D_\mu \Phi'^\dagger D^\mu \Phi'
- 
M^2 \Phi'^\dagger \Phi'
\notag \\
&&
-
\frac{C_0}{M^2}
F^2 \Phi'^\dagger \Phi'
+
\frac{C_2}{M^4}
T_{\mu \nu}
D^\mu \Phi'^\dagger
D^\nu \Phi'
,\end{eqnarray}
up to higher-order terms of mass-dimension ten. 
Notice that the operators with coefficients
$C_1$
and
$C_3$
vanish in uniform external fields, 
and are not required in our consideration.

The effect of the field redefinition Eq.~\eqref{eq:relFRD} on Green's functions happens to be innocuous. 
As in the toy-model example, 
we can compute the 
$\Phi$
propagator by first determining the 
$\Phi'$
propagator
$\cG'(x,y)$,
and then appealing to the field redefinition. 
Terms in the field redefinition involving  
$(D^2 + M^2)^n$
only produce contributions to the 
$\Phi$
two-point function 
$\cG(x,y)$
proportional to 
$\left( D^2 + M^2 \right)^{n-1}\delta(x-y)$. 
Such singular contributions have  
$\Phi$
and 
$\Phi^\dagger$
fields at the same spacetime point, 
and can be removed by imposing a renormalization condition on the vacuum energy. 
As a result, 
the 
$\Phi$
propagator has the form
\begin{eqnarray}
\cG(x,y)
&=& 
\left(
1 - \frac{\mathfrak{C}_2 - C_0 \mathfrak{C}_0 }{M^4} F^2
\right) 
\cG'(x,y)
,\end{eqnarray}
for 
$x_\mu \neq y_\mu$. 
Thus the two-point functions in the reduced and unreduced theories only differ by an overall constant. 
For on-shell states, 
the overall constant can be fixed by the wavefunction renormalization, 
i.e.~the residue at the pole. 
As expected, 
the field redefinition does not change the spectrum of the theory. 
In uniform electric fields, 
there is no on-shell condition for charged particles, 
however, 
the overall constant crucially does not alter the time dependence of the correlation function. 
In this way, 
the reduced and unreduced theories yield the same prediction for the behavior of the correlation function, 
which is of practical importance for lattice QCD analyses.%
\footnote{
In lattice QCD computations, 
moreover, 
the overall normalization of the two-point correlation function is unknown. 
With the contribution of the ground-state hadron isolated, 
the lattice correlator is proportional to the overlap factor between the chosen quark-level interpolating field and the ground-state hadron. 
}

Based on our analysis, 
we can use the theory specified by
Eq.~\eqref{eq:rel}
to describe the dynamics of a charged relativistic scalar coupled non-minimally to electromagnetism. 
While additional terms, 
such as those in 
Eq.~\eqref{eq:eomOs},
are needed to determine the Green's functions of the scalar propagating in external electromagnetic fields, 
these terms are not needed when we restrict our attention to uniform fields. 
In a uniform magnetic field, 
the correlation functions are unchanged provided that wavefunction renormalization has been accounted for in both reduced and unreduced theories. 
In a uniform electric field, 
which necessarily lacks the on-shell condition, 
the only modification to the two-point function is an overall constant.

\subsection{One- and Two-Photon Matching}

To discuss matching between relativistic and non-relativistic theories below, 
it is efficacious to relate the low-energy constants to observable quantities. 
The relativistic scalar hadron theory given in 
Eq.~\eqref{eq:rel}
depends on four unknown parameters, 
$C_0$--$C_3$. 
The operators with coefficients
$C_1$
and
$C_3$
obviously only contribute to processes involving at least one virtual photon. 
To relate these parameters to physical observables, 
we compute one- and two-photon processes. 
It is sufficient to treat processes with one virtual photon, 
and two real photons in order to determine all four parameters.

The scalar hadron's interaction with a virtual photon is described by the electromagnetic 
form factor, 
$F(q^2)$, 
entering current matrix elements between the scalar hadron. 
These matrix elements have the form
\begin{equation}
\langle \Phi (p') | J^\mu_{\text{e.m.}} | \Phi(p) \rangle
=
(p'+p)^\mu
F(q^2)
,\end{equation}
on account of gauge invariance and Lorentz covariance. 
In the small momentum transfer limit, 
we may expand the form factor to obtain
\begin{equation}
F(q^2) 
= 
Z 
+ 
\frac{1}{3!} q^2  < r^2>
+ 
\frac{1}{5!} q^4 < r^4>
+
\cdots
\label{eq:FF}
.\end{equation}
The form factor at vanishing momentum-transfer is constrained by the Ward identity to be the total charge. 
The first-order correction is conventionally parameterized by defining a charge radius
$\sqrt{< r^2 >}$, 
and the second-order correction we define as being a higher moment of the charge distribution, 
$< r^4>$. 
The physical interpretation of both of these quantities is complicated by relativistic effects, 
however, 
one can identify them as moments of the transverse distribution of charge in the infinite momentum frame~\cite{Burkardt:2002hr}. 
Deriving the electromagnetic current from the relativistic action enables us to compute the scalar hadron's form factor. 
In doing so, 
we find the simple relations
\begin{eqnarray}
\frac{C_1}{M^2} 
&=& 
\frac{1}{3!}
< r^2>
,
\quad
\frac{C_3}{M^4} 
= 
\frac{1}{5!}
< r^4>
.\end{eqnarray}

The remaining two coefficients can be related to physical observables by considering two-photon processes. 
To this end, 
we consider the real Compton scattering process, 
$\gamma(k) + \Phi(p) \to \gamma(k') + \Phi(p')$. 
Working in the laboratory frame, 
the forward and backward Compton amplitudes are given in a low-energy expansion by%
~\cite{Schumacher:2005an}
\begin{eqnarray}
T(\theta = 0)
&=&
\vec{\varepsilon} \, {}^{\prime *} \cdot \vec{\varepsilon}
\left[ - \frac{Z^2}{M} +  4 \pi (\alpha_E + \beta_M) \omega^2 \right],
\notag \\
T(\theta = \pi)
&=&
\vec{\varepsilon} \, {}^{\prime *} \cdot \vec{\varepsilon}
\left[ - \frac{Z^2}{M} + 4 \pi (\alpha_E - \beta_M) \omega \omega' \right]
,\end{eqnarray}
where 
$\alpha_E$
and
$\beta_M$
are the electric and magnetic polarizabilities,
respectively. 
Using the theory defined by Eq.~\eqref{eq:rel} to compute the Compton amplitude at small photon energy determines the values,
\begin{eqnarray}
C_0
&=&
\pi M^3 ( \alpha_E - \beta_M),
\quad
C_2
=
4\pi M^3 (\alpha_E + \beta_M)
.\end{eqnarray}
Physically the Compton scattering cross section can be written as the coherent sum of contributions from photon helicity preserving, 
$\Delta \lambda = 0$,
and helicity flip,
$\Delta \lambda = 2$, 
processes. 
The operator with coefficient 
$C_0$ 
contributes exclusively to the former, 
while the operator with coefficient 
$C_2$
contributes exclusively to the latter.

\subsection{Uniform External Fields}
\label{s:corr_QED}

To further aid in matching relativistic and non-relativistic theories below, 
we investigate the charged particle correlation functions in external magnetic and electric fields. 
These external fields are chosen to be uniform; 
and, 
because the correlation functions depend on the gauge, 
particular gauges are employed. 
It is straightforward to implement different gauge choices.

\subsubsection{Magnetic Field}

A charged particle propagating in a uniform magnetic field can be projected onto states of definite energy. 
For this case, 
we choose to align the magnetic field with the 
$z$-direction, 
and accordingly choose the gauge potential 
$A_\mu = - B x_2 \, \delta_{\mu 1}$. 
The 
$\Phi$
propagator has an infinite number of poles corresponding to the various Landau levels. 
This feature is best exhibited by employing 
Schwinger's proper-time trick~\cite{Schwinger:1951nm}.
The coordinate wavefunction of the 
$n$-th Landau level, 
$\psi_n(x_2)$, 
allows us to project out this energy eigenstate due to orthogonality%
~\cite{Tiburzi:2012ks}. 
This can be seen, 
for example, 
by computing the propagator projected at the sink
\begin{equation}
\cG_B^{(n)} (t,0)
= 
\int d \vx  \, \psi^*_n(x_2) 
\langle 0|\Phi (\vx,t) \Phi^\dagger (\vzero,0)|0\rangle_B
,\end{equation}
which has the simple behavior
$\cG_B^{(n)} (t,0) = Z_n e^{ - i E_n t}$, 
assuming that
$t>0$. 
The energy 
$E_n$ 
is given by
\begin{equation} \label{eq:landau}
E_n
= 
\Big[
M^2 + | Z B| ( 2 n + 1) - 4 \pi \beta_M M B^2
\Big]^{1/2}
,\end{equation}
where we have replaced 
$\frac{1}{4 M^2} \left( C_2 - 4 C_0 \right) =  4\pi \beta_M M$
to express the energy in terms of the magnetic polarizability. 
The expansion of the energy in powers of 
$M$
enables straightforward comparison with the non-relativistic theory.

\subsubsection{Electric Field}

For a charged particle in a uniform electric field,
however, 
the situation is more involved. 
Specifying the four-vector potential 
$A_\mu = - E t \, \delta_{\mu 3}$, 
and rescaling the 
$\Phi$
field leads to the Lagrange density 
\begin{equation}
\cL
=
\Phi^\dagger_{\vec{p} = 0}
\left[
- \frac{\partial^2}{\partial t^2}
- (Z E t)^2
- M^2 
+ 4 \pi \alpha_E M E^2
\right]
\Phi_{\vec{p} = 0}
\label{eq:mess}
,\end{equation}
where we have dropped terms of order 
$E^4$, 
and projected the field onto vanishing three momentum
$\vec{p} = 0$.
We have also rewritten the combination of low-energy parameters, 
$\frac{1}{4 M^2} \left(C_2 + 4 C_0\right) = 4 \pi \alpha_E M$, 
in favor of the electric polarizability. 
The propagator resulting from 
Eq.~\eqref{eq:mess}
contains singularities associated with the real-time production of any number of particle-antiparticle pairs. 
This is the Schwinger mechanism.

To avoid the Schwinger mechanism altogether, 
we work in Euclidean space.  
This choice is further motivated by lattice QCD computations which are necessarily carried out in Euclidean space. 
With 
$t = - i \tau$, 
and
$E = i \cE$, 
we have the Euclidean action density
\begin{equation}
\cL_E
=
\Phi^\dagger_{\vec{p} = 0}
\left[
- \frac{\partial^2}{\partial \tau^2}
+ (Z \cE \tau)^2
+M^2 
+ 4 \pi \alpha_E M \cE^2
\right]
\Phi_{\vec{p} = 0}
\label{eq:messE}
.\end{equation}
From this action density, 
one can compute the two-point function
\beq
\cG_{\cE}(\tau,0)
=
\int d\vx  \,
\langle 0|\Phi (\vx,\tau)
\Phi^\dagger (\vzero,0)|0\rangle_{\cE},
\eeq
where 
$\tau > 0$
is implicitly assumed. 
Appealing again to Schwinger's proper-time trick, 
we find an integral representation for the propagator~\cite{Tiburzi:2008ma}
\beq
\cG_\cE(\tau,0)
=
\frac{1}{2}
\int_0^\infty ds
\sqrt{\frac{Z\cE}{2\pi \sinh Z\cE s}}
e^{
-
\frac{1}{2}
\left(
\frac{Z\cE \tau^2}{\tanh Z\cE s}
+
E_\cE^2 s
\right)
},\nn \\
\label{eq:rel_prop_E}
\eeq
where the quantity 
$E_\cE$
can roughly be termed the relativistic initial energy, 
\emph{cf}.~the behavior of Eq.~\eqref{eq:messE} at $\tau = 0$, 
and is given by  
\beq
E_\cE
=
\left[
M^2 + 4\pi \alpha_E M \cE^2
\right]^{1/2} 
.\eeq
Notice that the electric polarizability produces a positive shift of the initial energy in Euclidean space. 

Due to the lack of energy eigenstates, 
the long-time behavior of the correlator in 
Eq.~\eqref{eq:rel_prop_E} 
does not exhibit the exponential decay that is characteristic of correlation functions in Euclidean space. 
The logarithmic derivative of the correlator grows in Euclidean time, 
which roughly corresponds to the particle acquiring energy from the electric field. 
Unfortunately the proper-time integration cannot be performed in closed form. 
The non-relativistic reduction of this propagator will be carried out, 
and compared with the propagator computed from the non-relativistic effective theory.

\section{Non-Relativistic Scalar QED} \label{s:NRscalar} %

For sufficiently low energies, 
one can formulate the effective theory of a charged composite scalar using 
HQET 
power counting. 
This theory is organized in inverse powers of the particle's mass 
$M$, 
which is treated as a large energy scale, 
see~\cite{Manohar:2000dt}. 
In considering the dynamics of a charged scalar in external electromagnetic fields, 
we must address the effects of operators related by the non-relativistic equations of motion. 
In a uniform electric field, 
we find the non-relativistic effective theory requires an additional such operator.

\subsection{Action and Relativistic Invariance}   %

To write down the non-relativistic theory, 
we consider the most general Lagrange density for a charged composite scalar 
$\phi$ 
interacting with electromagnetic fields. 
We impose Hermiticity,
and
invariance under 
$P$, 
$T$, 
and gauge transformations.
Including all terms up to 
$\cO(M^{-4})$, 
we find
\beq
\cL&=&\phi^\dagger\left[
iD_0
+
c_2 \frac{\vD^2}{2M}
+
c_D \frac{[\vDel\cdot\vE]}{8M^2}
\right.\nn \\
&&
+
c_4\frac{\vD^4}{8M^3}\left.
+
ic_M
\frac{\{D^i,[\vDel\times\vB]\}}{8M^3}
+
c_{A_1}
\frac{\vB^2-\vE^2}{8M^3}
\right.\nn\\
&&
\left.
-
c_{A_2}\frac{\vE^2}{16M^3}
+
c_{X_0}\frac{[iD_0,\vD\cdot\vE+\vE\cdot\vD]}{8M^3}
\right.\nn\\
&&
\left.
+
c_{X_1}
\frac{[\vD^2,\vD\cdot\vE+\vE\cdot\vD]}{16M^4}
+
c_{X_2}\frac{\{\vD^2,[\vDel\cdot\vE]\}}{16M^4}
\right.\nn\\
&&
\left.
+
c_{X_3}\frac{[\vDel^2\vDel\cdot\vE]}{16M^4}
+
ic_{X_4}\frac{\{D^i,(\vE\times\vB)^i\}}{16M^4}
\right]\phi.
\label{eq:nonrel}
\eeq
In the non-relativistic theory,
the gauge covariant derivative is specified by 
\beq
D_0 &=& \partial_0+iZA_0, 
\notag
\\  
D^i &=& \nabla^i-iZA^i
.\eeq 
The electric and magnetic fields 
$\vE$ 
and 
$\vB$
are given by standard expressions,
$\vE=-\partial_0\vA-\vec{\nabla}A_0$ 
and 
$\vB=\vec{\nabla}\times\vA$, 
respectively. 
It will be useful in what follows to recall the commutators of two covariant derivatives, 
$[D^i,D^j]=
-iZ\epsilon^{ijk}B^k$ 
and 
$[D^i,D_0]=-iZE^i$. 
Note that the product $i\vB$ is time-reversal even and the factor $i$ 
necessitates the anti-commutator structure to satisfy Hermiticity. 
The other time-reversal even quantity involving the magnetic field can be 
eliminated by using Maxwell's equation, 
$\partial_0\vB=-\vec{\nabla}\times \vE$. 
Anti-commutator terms with two derivatives are Hermitian without a factor of $i$, 
while $c_{X_0}$ and $c_{X_1}$ terms require a commutator for Hermiticity.

Not all of the operator coefficients appearing in the effective theory are independent parameters, 
because Lorentz invariance implies relations between different orders in the 
$1/M$
expansion~\cite{Luke:1992cs,Heinonen:2012km}. 
Such relations can be deduced by performing an infinitesimal boost, 
and demanding invariance order-by-order in 
$1/M$. 
This is the variational method detailed in~\cite{Hill:2012rh}. 
Parameterizing the boost with momentum 
$\vec{q}$, 
we have the variations
\beq
\delta \vec{D} = \vec{q} \, D_0/M,
\quad
\delta D_0 = \vec{q} \cdot \vec{D} / M
,\notag\eeq 
along with
\beq
\delta \vec{B} = - \vec{q} \times \vec{E} / M, 
\quad
\delta \vec{E} = \vec{q} \times \vec{B} / M
.\notag\eeq 
We also require the transformation property of the scalar field, 
which can be written to 
$\cO(M^{-4})$
 in the form
\beq
\phi(x)
&\to&
e^{ - i \vec{q} \cdot \vec{x}}
\left[
1 
+ 
A \frac{i \vec{q} \cdot \vec{D}}{2 M^2}
+ 
B \frac{ i \vec{q} \cdot \vec{E}}{8 M^3}
\right.
\nn \\
&&
\left.
+
C \frac{\big\{ i \vec{q} \cdot \vec{D}, \vec{D} {}^2 \big\}}{8 M^4}
+
D \frac{\vec{q} \cdot [ \vec{\nabla} \times \vec{B}]}{8 M^4}
\right.
\nn \\
&&
\left.
+ 
E \frac{\epsilon^{ijk} q^i \left\{ B^j, D^k \right\}}{8 M^4}
+
F \frac{\big\{ D_0,  \vec{q} \cdot \vec{E} \big\}}{8 M^4}
\right]
\phi(x)
,
\nn \\
\eeq
where the parameters
$A$--$F$
remain to be determined.

Boost invariance can be enforced order-by-order in 
$1/M$
provided the field transformation is specified by the parameters
\begin{align}
A = 1,
\qquad
B = c_D + 2 c_{X_0}, 
\qquad
C = 1, 
\nn \\
D = - c_M,
\quad
E =  - c_{X_1},
\quad
F = c_{X_0}
.\end{align}
Furthermore, 
the coefficients of operators in the effective theory must satisfy the relations
\begin{eqnarray}
c_2 = c_4 &=& 1,
\quad
c_M = \frac{1}{2} c_D,
\quad
c_{X_1} - c_{X_0} = \frac{1}{2} ( Z + c_D),
\nn \\
c_{X_2} &=& 0,
\quad
c_{X_4} = 2 Z c_D - c_{A_2}
\label{eq:relations}
.\end{eqnarray} 
Taking into account these relations, 
there are five unconstrained parameters in the effective theory.

With the exception of the operator having coefficient 
$c_{X_0}$, 
the operators enumerated in 
Eq.~\eqref{eq:nonrel}
are identical to the spin-independent operators found in~\cite{Hill:2012rh}. 
For on-shell processes involving 
$\phi$, 
we furthermore have the operator equivalence
\begin{equation}
\phi^\dagger
\frac{[iD_0,\vD\cdot\vE+\vE\cdot\vD]}{8M^3}
\phi
\overset{\text{eom}}{=}
-
\phi^\dagger
\frac{[\vec{D} {}^2, \vD\cdot\vE+\vE\cdot\vD]}{ 16 M^4}
\phi
\label{eq:eom}
,\end{equation}
which arises from applying the HQET equations of motion. 
Consequently, 
there are only four independent parameters required to describe on-shell process. 
As we will see, 
however,
the remaining parameter
$c_{X_0}$
is necessary to describe the Green's functions in a uniform electric field.

Notice we did not write down all possible operators related by the equations of motion. 
For example,  
operators of the form
$\phi^\dagger (i D_0)^n \phi$
for 
$n > 1$
have been excluded. 
The difference between these operators and their counterparts related by the equations of motion
is a time-dependent modification of Green's functions
$G(t',t)$
by singluar terms involving derivatives of delta functions,
$\left(
D_0
\right)^{n-1} \delta(t' - t)$. 
For this reason, 
these operators have been excluded. 
Beyond such operators, 
there are further terms, 
for example the operator
$\vE^2 \phi^\dagger  i D_0 \phi$, 
which can modify the time dependence of Green's functions in a non-constant electric field. 
We have omitted this operator, 
however, 
because it only modifies Green's functions by an overall constant in uniform electric fields. 
In writing 
Eq.~\eqref{eq:nonrel}, 
we are claiming that the operator with coefficient 
$c_{X_0}$
is the only operator related by the equations of motion that is required to address the case of uniform electromagnetic fields
at 
$\cO(M^{-3})$. 
The appearance of additional equation-of-motion operators for uniform fields at 
$\cO(M^{-4})$
has not been considered.

\subsection{One- and Two-Photon Matching}

To determine the phenomenological values of the non-relativistic effective field theory coefficients, 
we perform one- and two-photon matching similar to that carried out above in the relativistic case. 
This is the scalar analogue of non-relativistic effective field theory matching carried out in~%
\cite{Manohar:1997qy,Hill:2012rh}.
The resulting matching conditions will additionally establish the relations between relativistic and non-relativistic
low-energy constants. 
As above, 
we restrict our attention to processes involving either one virtual photon, or two real photons.

For virtual photon scattering with the 
$\phi$, 
we use the relativistic decomposition of the form factor given in Eq.~\eqref{eq:FF}. 
Kinematically,
the momentum transfer squared has the non-relativistic expansion
\begin{equation}
q^2 
= 
- \vec{q} \, {}^2
+ 
\frac{1}{4 M^2}
\left( \vec{q} \, {}^2 + 2 \vec{q} \cdot \vec{p} \, \right)
+ 
\cdots
.\end{equation}
The matrix element of the charge density operator in turn has the non-relativistic expansion
\beq
\langle 
\vec{p} \, ' | J^0 | \vec{p} \, \rangle
&=&
Z
- 
\frac{\vec{q} \, {}^2}{3!}
<r^2>
+ 
\frac{\vec{q} \, {}^4}{5!}
<r^4>
\nn \\
&&
+
\frac{( \vec{q} \, {}^2 + 2 \vec{q} \cdot \vec{p} \, )^2}{16 M^4}
\left[ \frac{Z}{2}  + \frac{2}{3} M^2 < r^2> \right]
, \ \ \eeq
in an arbitrary frame. 
In the above expression, 
we have accounted for the differing normalization between relativistic and non-relativistic states,
see Eq.~\eqref{eq:Phi_vs_phi} below. 
Computation of the same matrix element using the HQET action in Eq.~\eqref{eq:nonrel}, 
produces the relations
\begin{eqnarray}
c_D
= \frac{4}{3} M^2 <r^2>
, 
\quad
c_{X_3} = \frac{2}{15} M^4  <r^4>,
\nn \\
c_{X_1} - c_{X_0}
= 
\frac{1}{2} \left( Z + \frac{4}{3} M^2 < r^2> \right)
,
\quad
c_{X_2} 
= 
0
.  \ \ \end{eqnarray}
The latter two relations are required by the imposition of Lorentz invariance, 
see Eq.~\eqref{eq:relations}. 
Matching the spatial current matrix element in an arbitrary frame confirms the relation
$c_M = \frac{1}{2} c_D$.

Evaluation of the real Compton scattering amplitude is simplified in the non-relativistic limit. 
In the laboratory frame, 
the final-state photon frequency satisfies the condition
$\omega' = \omega + \cO(\omega / M)$. 
Computing the Compton amplitude up to 
$\cO(M^{-3})$
accuracy, 
we determine the matching conditions
\beq
16 \pi M^3 \alpha_E
&=&
Z c_D - c_{A_1} - \frac{1}{2} c_{A_2},
\quad 
16 \pi M^3 \beta_M
=
c_{A_1}
,
\nn \\
\label{eq:twomatch}
\eeq
which relate low-energy constants to the electric and magnetic polarizabilities.

From one- and two- photon processes, 
we have thus determined the four on-shell parameters of the effective theory in terms of physical observables. 
The parameter
$c_{X_0}$
cannot be determined in this way, 
because physical processes only depend on the linear combination 
$c_{X_1} - c_{X_0}$, 
\emph{cf}.~Eq.~\eqref{eq:eom}.
Comparing the matching conditions between relativistic and non-relativistic theories 
enables us to relate the low-energy constants of the two effective theories. 
From single-photon matching, 
we find the relations
\beq
c_D 
=
8 C_1
,\quad 
c_{X_3}
=
16 C_{3}
,\eeq
which shows that these low-energy constants are determined entirely from 
virtual photon couplings in the relativistic theory. 
The two-photon matching conditions produce the relations
\beq
c_{A_1} 
= 
2 (C_2 - 4 C_0), 
\quad
c_{A_2}
=
8( 2 Z C_1 - C_2)
.\eeq
Notice the parameter
$c_{A_2}$
has a piece
$\propto C_1$ 
that arises from a relativistic operator contributing exclusively to virtual photon processes. 
This produces exact cancelation of the 
$c_D$ 
term contributing to 
$\alpha_E$ 
in Eq.~\eqref{eq:twomatch}, 
which is required because 
$\alpha_E$
can be determined from Compton scattering with two 
\emph{real} 
photons.

The remaining on-shell parameters of the non-relativistic theory are constrained by Lorentz invariance. 
For completeness, 
the remaining relations between non-relativistic and relativistic low-energy constants are
\beq
c_M = 4 C_1
,\quad
c_{X_1} - c_{X_0}
= 
\frac{1}{2} ( Z + 8 C_1)
,\quad
c_{X_4} = 8 C_2
.\nn \\
\eeq
As far as on-shell processes are concerned, 
we can employ the effective theory in 
Eq.~\eqref{eq:nonrel}
omitting the operator with coefficient
$c_{X_0}$. 
This is not the case when we consider the Green's functions in a uniform electric field. 
We now turn our attention to background electromagnetic fields.

\subsection{Uniform External Fields}
\label{s:corr_NRQED}

We consider the non-relativistic effective theory in background electromagnetic fields. 
First we show that there are no complications for the case of a uniform magnetic field. 
For a uniform electric field, 
we expose the difficulty of dropping the operator with coefficient 
$c_{X_0}$. 
We then determine this coefficient by matching Green's functions calculated with 
HQET 
power counting 
and the corresponding 
HQET 
expansion of the relativistic Green's function. 
The matching condition for 
$c_{X_0}$
is verified by repeating the Green's function matching with 
NRQED power counting.

\subsubsection{Magnetic Field}

To match Green's functions, 
let us first consider the case of a uniform magnetic field specified, 
as above, 
by the vector potential
$\vec{A} = - x_2 B \hat{x}_1$. 
In such an external field,  
the HQET action reduces to
\beq
\cL&=&\phi^\dagger_{\vec{p}_\perp = 0}
\left[
i \partial_0
- 
H
+
\frac{H^2}{2M^2}
+
c_{A_1}
\frac{B^2}{8M^3}
\right] 
\phi_{\vec{p}_\perp = 0}
,\label{eq:Bfield}
\eeq
up to terms of order
$M^{-5}$. 
Notice we have projected onto the sector of zero transverse momentum 
$\vec{p}_\perp = (p_1, p_3) = 0$,
for simplicity;
and, 
$H$
is the harmonic oscillator Hamiltonian given by
$H = \frac{1}{2M} \left[ - \partial_2^2 + (ZB)^2 x_2^2 \right]$. 
Expanding in the oscillator basis, 
we see the energy eigenvalues have the form
\beq
E_n^{\text{NR}}
=
\frac{|Z B|}{M} (n+\frac{1}{2})
-
\frac{(Z B)^2}{2 M^3} (n+ \frac{1}{2})^2
-
\frac{1}{2} 4 \pi \beta_M B^2
,\nn \\
\eeq
having traded the low-energy constant 
$c_{A_1}$
for the magnetic polarizability 
$\beta_M$
through the matching condition, 
Eq.~\eqref{eq:twomatch}. 
Comparing with the relativistic spectrum from 
Eq.~\eqref{eq:landau}, 
we see they agree
\beq
E_n - M 
= 
E_n^{\text{NR}}
+ 
\cO(M^{-5})
,\eeq
to the order we are working in the HQET expansion. 
Because the single-particle wave-functions of the Landau levels also agree, 
the two-point correlation function calculated in HQET matches with the non-relativistic expansion of the relativistic correlation function. 
As expected, 
no operators related by equations of motion are required for this case.

\subsubsection{Electric Field: HQET}
\label{s:EHQET}

Turning our attention to the case of a uniform electric field, 
we first write the non-relativistic action in Euclidean space. 
Specifying a uniform electric field through the vector potential
$\vec{A} = - \cE \tau \hat{x}_3$, 
as above, 
we have the HQET action density
\beq
\cL_E
&=&
\phi^\dagger_{\vec{p} = 0}
\left[
\frac{\partial}{\partial \tau}
+
\frac{( Z \cE \tau)^2}{2M}
-
\frac{( Z \cE \tau)^4}{8M^3}
+
\frac{ c_\text{NR} \, \cE^2}{16 M^3}
\right]\phi_{\vec{p} = 0},
\nn \\
\label{eq:HQET_E}
\eeq 
having projected onto the sector of vanishing three-momentum for ease. 
Above,
we employ the abbreviation
\begin{equation}
c_\text{NR}
= 
- 
2 c_{A_1} 
- 
c_{A_2} 
-
4 Z c_{X_0}
\label{eq:CNR}
,\end{equation}
for the linear combination of low-energy constants multiplying the electric-field-squared operator. 
Corrections to this action are of order 
$M^{-5}$.
The coefficient 
$c_\text{NR}$ 
is surprising for two reasons. 
First it depends on the linear combination of low-energy parameters 
$ - c_{A_1} - \frac{1}{2} c_{A_2} = 16 \pi M^3 \alpha_E - Z c_D$, 
where the left-hand side of the equation makes use of the matching condition in 
Eq.~\eqref{eq:twomatch}. 
The 
$\cE^2$
shift of the action depends on the electric polarizability 
$\alpha_E$
as it must, 
however, 
there is also a contribution from the charge radius, 
$c_D$,
which is physically impossible because the external field cannot probe virtual photon couplings.%
\footnote{
Without the equation-of-motion operator, 
we would set its coefficient to zero, 
$c_{X_0} = 0$,
and accordingly the shift of the initial energy arising from 
$c_\text{NR}$
in 
Eq.~\eqref{eq:HQET_E} appears in Minkowski space exactly as shown in Eq.~\eqref{eq:one}. 
} 
The second surprise is the appearance of the contribution proportional to 
$c_{X_0}$
which arises from the operator related by the equations of motion. 
To obtain the correct physics, 
the first surprising feature of 
Eq.~\eqref{eq:CNR}
actually requires the second surprising feature for cancellation of the offending term.

Let us further scrutinize the appearance of the coupling 
$c_{X_0}$
in 
Eq.~\eqref{eq:CNR}. 
Notice the equation of motion equivalence shown in 
Eq.~\eqref{eq:eom}
becomes invalid in a uniform electric field due to the striking difference in evaluating the two terms
\beq
\left[
iD_0, 
\vD\cdot\vE
+
\vE\cdot\vD
\right]
&=&
- 2 Z \vE {}^2 
,\nn \\
\left[
\vec{D} {}^2, 
\vD\cdot\vE
+
\vE\cdot\vD
\right]
&=&
0
\label{eq:striking}
.\eeq
Because the latter operator vanishes in a uniform electric field, 
the corresponding coupling 
$c_{X_1}$
disappears from the action. 
Consequently the Green's function does not depend on the linear combination 
$c_{X_1} - c_{X_0}$ 
which enters on-shell processes. 
Instead it depends on the parameter 
$c_{X_0}$, 
which we have yet to determine.

The fact that the equation-of-motion-related operator is relevant to the uniform electric field case is 
further evidenced by considering the field redefinition that can be employed for its removal. 
To remove the operator with coefficient
$c_{X_0}$
from the HQET action, 
Eq.~\eqref{eq:nonrel}, 
we invoke the field redefinition
\beq
\phi 
= 
\left( 
1 - c_{X_0} \frac{\vD \cdot \vE + \vE \cdot \vD}{8 M^3}
\right)
\phi'
\label{eq:FRDNR}
.\eeq
Rewritten in terms of the 
$\phi'$
field, 
the equation-of-motion operator has been removed, 
and the related operator now has coefficient
$c_{X_1} - c_{X_0}$. 
In a uniform electric field, 
the related operator vanishes by Eq.~\eqref{eq:striking}. 
The Green's functions for the fields 
$\phi$
and 
$\phi'$, 
however, 
are different due to the field redefinition employed in Eq.~\eqref{eq:FRDNR}. 
In particular, 
the Euclidean time correlation functions in a uniform electric field
\beq
G_\cE (\tau,0)
&=&
\int d\vec{x}  \,
\langle 0 | \phi(\vec{x}, \tau) \phi^\dagger (0,0) | 0 \rangle_\cE
,\nn \\
G'_\cE (\tau,0)
&=&
\int d\vec{x}  \,
\langle 0 | \phi'(\vec{x}, \tau) \phi'^\dagger (0,0) | 0 \rangle_\cE
,\eeq
are related by
\begin{equation}
G_\cE(\tau,0)
=
\left( 
1 
+
c_{X_0} \frac{Z \cE^2 \tau}{4 M^3}
\right)
G_\cE'(\tau,0)
\label{eq:relationism}
.\end{equation}
Hence the correlation functions have visibly different time dependence. 
This difference between correlation functions, 
moreover, 
can be alternately obtained by treating the 
$c_{X_0}$
term present in the action, 
Eq.~\eqref{eq:HQET_E},
in perturbation theory.

Having argued that the 
$c_{X_0}$
term belongs in the HQET action for a uniform electric field,
we must determine this parameter. 
A way to determine 
$c_{X_0}$
is to start with the fully relativistic scalar propagator in an electric field, 
and perform the HQET expansion. 
Matching the behavior of the propagator order-by-order in 
$M^{-1}$
will yield the value of this parameter. 
Thought of in this way, 
the external electric field problem requires an additional matching relation due to the lack of an 
on-shell condition. 
The coefficient 
$c_{X_0}$, 
which cannot be resolved from on-shell processes, 
can be determined at the level of Green's functions.

Computing the Euclidean two-point correlation function for 
$\phi$
directly from 
Eq.~\eqref{eq:HQET_E}, 
we arrive at
\beq
G_\cE(\tau,0)
&=&
\theta(\tau)
\Bigg[
1
- 
\frac{(Z \cE)^2 \tau^3}{6 M}
+
\frac{(Z \cE)^4 \tau^6}{72 M^2}
\nn \\
&&
+ 
\frac{(Z \cE)^4 \tau^5}{40 M^3}
-
\frac{(Z \cE)^6 \tau^9}{1296 M^3}
- 
c_{\text{NR}}
\frac{\cE^2 \tau}{16 M^3}
\Bigg]
.\eeq
On the other hand, 
carrying out the 
$1/M$
expansion of the relativistic correlation function 
$\cG_\cE(\tau,0)$
in Eq.~\eqref{eq:rel_prop_E}, 
and appropriately accounting for the difference in normalization factors
[see Eq.~\eqref{eq:Phi_vs_phi} below], 
we find the difference between relativistic and non-relativistic correlators is given by
\beq
\Delta G_\cE (\tau,0) 
= 
\theta(\tau) 
\frac{\cE^2 \tau}{16 M^3}
\left( c_{\text{NR}} - c_\text{R} \right)
,\eeq
where the coefficient 
$c_\text{R}$
arises from the relativistic correlation function, 
and is given by
\begin{equation}
c_\text{R} = 32 \pi M^3 \alpha_E + 4 Z^2
.\end{equation} 
This coefficient produces a perturbative correction to the non-relativistic initial energy having the form
$\Delta E = - \frac{1}{2} \left( 4 \pi \alpha_E + \frac{Z^2}{2 M^3} \right) \vE^2 $,
in Minkowski space. 
This result is to be contrasted with that in Eq.~\eqref{eq:one}, which was obtained by incorrectly dropping the equation-of-motion operator.

Requiring that the correlation functions match 
demands that 
$c_\text{NR} = c_\text{R}$,
and
allows us to determine 
\beq \label{eq:missing}
c_{X_0} 
= 
-
\frac{1}{2} c_D 
-
Z 
.\eeq
Having determined this final parameter, 
the time-dependence of the HQET propagator in a uniform electric field is fully specified. 
In practice, 
the HQET expansion is insufficient to describe lattice QCD data. 
While the external electric field may be weak, 
large corrections will arise in the long-time limit of the correlator. 
To this end, 
it is efficacious to include the Euclidean time 
$\tau$
in the power counting, 
and thus we turn to NRQED.

\subsubsection{Electric Field: NRQED}  %
\label{s:ENRQED}

The parameter 
$c_{X_0}$
can also be determined from carrying out the matching of correlation functions using 
NRQED power counting. 
This power counting, 
moreover, 
leads to a useful expansion of the relativistic correlation function that potentially could  
simplify the analysis of lattice QCD data. 
HQET and NRQCD effective theories share the same Lagrange density, 
however, 
the ordering of operators is different. 
Instead of counting powers of 
$1/M$, 
the NRQCD counting is organized in powers of the small velocity 
$v$~\cite{Luke:1999kz}.

For a charged particle in a uniform electromagnetic field, 
we employ NRQED power counting in which 
$D_0$ 
and 
$\vD^2$ 
both count as 
$\cO(v^2)$. 
Consequently explicit factors of the time 
$t$ 
count as 
$\cO(v^{-2})$, 
while the electric and magnetic fields,
$\vE$
and
$\vB$,
count as 
$\cO(v^3)$. 
Keeping all terms of the HQET Lagrange density in uniform electromagnetic fields up to order 
$v^6$, 
we have
\beq
\cL
&=&
\phi^\dagger
\left[
iD_0+\frac{\vD^2}{2M}
+
\frac{\vD^4}{8M^3}
+
\frac{\vD^6}{16M^5}
\right. \nn \\
&&\left.
+
c_\text{NR}
\frac{\vE^2}{16M^3}
+
c_{A_1}
\frac{\vB^2}{8M^3}
\right]
\phi.
\label{eq:nonrelQED}
\eeq
Notice at this order, 
a further term from the HQET Lagrange density is required. 
This term is the next-order relativistic correction to the kinetic energy, 
and is the only term at 
$\cO(M^{-5})$. 
In NRQED, 
this term contributes at 
$\cO(v^6)$
which is the same order required to determine the electric and magnetic polarizabilities.

To carry out the matching, 
we work in Euclidean space. 
For a uniform electric field, 
the Euclidean NRQED action density is given by
\beq
\cL_E
&=&
\phi^\dagger_{\vec{p} = 0}
\left[
\frac{\partial}{\partial\tau}
+
\frac{(Z\cE\tau)^2}{2M}
-
\frac{(Z\cE\tau)^4}{8M^3}
\right.
\nn \\
&&\left.
+
\frac{(Z\cE\tau)^6}{16M^5}
+ 
c_\text{NR} \frac{\cE^2}{16 M^3}
\right]
\phi_{\vec{p} =0}
\label{eq:nonrel_uniform_E}
,\eeq
in the sector of vanishing three-momentum. 
Because the action involves only a first-order differential operator, 
we can easily determine the Green's function 
\beq
G_\cE(\tau,0)
&=&
\theta(\tau) 
\,
\textrm{exp}
\left[
-
\frac{(Z\cE)^2\tau^3}{6M}
+
\frac{(Z\cE)^4\tau^5}{40M^3}
\right.\nn\\
&&\left.
-
\frac{(Z\cE)^6\tau^7}{112M^5}
- 
c_\text{NR}
\frac{\cE^2\tau}{16 M^3}
\right].
\label{eq:nonrel_prop_E}
\eeq
Notice that the first term in the exponential is 
$\cO(v^0)$, 
while the second term is 
$\cO(v^2)$, 
and the last two terms are both 
$\cO(v^4)$. 
These latter terms need not be exponentiated, 
but can be expanded out to 
$\cO(v^4)$. 
The 
$\cO(v^0)$
term was derived in the original proposal for treating charged hadrons in external fields%
~\cite{Detmold:2006vu}.
The present result provides a useful extension including relativistic corrections in a systematic way.

The NRQED expansion of the relativistic propagator 
$\cG_\cE(\tau,0)$
in Eq.~\eqref{eq:rel_prop_E}
requires 
Laplace's method. 
The technical details of the expansion are presented in Appendix~\ref{s:A}. 
Up to 
$\cO(v^4)$, 
we obtain the same form of the correlation function as in 
Eq.~\eqref{eq:nonrel_prop_E}
with the exception that the coefficient
$c_\text{NR}$ 
is replaced by 
$c_\text{R}$. 
Matching the NRQED correlators then requires
$c_\text{NR} = c_\text{R}$, 
which consequently leads to the value of 
$c_{X_0}$
obtained in Eq.~\eqref{eq:missing} above. 
We have established that NRQED matching of the Green's functions yields the same result.

\subsection{Non-Relativistic Expansion of Relativistic QED}%
\label{s:NRRQED}

As a final check of our results, 
we determine the parameter
$c_{X_0}$
using a brute-force expansion of the relativistic Lagrange density 
with a careful treatment of the equations of motion. 
While such an expansion is rather antithetical to the effective field theory mindset, 
it can be carried out straightforwardly for a scalar particle, 
and the non-relativistic matching can thus be performed directly at the level of the action. 
We consider the expansion to 
$\cO(M^{-3})$.

To reduce the relativistic theory in Eq.~\eqref{eq:rel} to the non-relativistic theory in Eq.~\eqref{eq:nonrel}, 
we need the relation between the relativistic scalar field 
$\Phi$ 
and non-relativistic scalar field 
$\phi$, 
which is 
\beq
\Phi(x)
=
\frac{e^{-iMt}}{[4(M^2-\vD^2)]^{1/4}}
\phi(x).
\label{eq:Phi_vs_phi}
\eeq
This relation has already been used throughout to convert between the relativistic and non-relativistic normalization of states.

After inserting the relation between relativistic and non-relativistic fields into the relativistic action,  Eq.~\eqref{eq:rel}, 
we perform the 
$1/M$-expansion keeping all terms up to 
$\cO(M^{-3})$.
Many of the terms in Eq.~\eqref{eq:nonrel} automatically appear in the expansion, 
however, 
there are also additional terms. 
Explicitly, 
we have 
\beq 
\cL 
&=&
\phi^\dagger 
\Bigg[
i D_0 
+ 
\frac{\vD^2}{2M}
- 
\frac{D_0^2}{2M}
+
\frac{\big\{ i D_0, \vD^2 \big\}}{4 M^2}
+ 
c_D
\frac{[ \vDel \cdot \vE]}{8 M^2} 
\nn \\
&&
+ 
\frac{\vD^4}{4 M^3}
- 
\frac{\big\{ D_0^2, \vD^2 \big\}}{8 M^3}
+
c_{A_1}
\frac{\vB^2 - \vE^2 }{8 M^3}
-
c_{A_2}
\frac{\vE^2}{16 M^3}
\nn \\
&&
+
\frac{i c_D}{16 M^3}
\big\{
D^i, 
[ \vDel \times \vB ]^i
\big\}
+ 
\frac{c_D}{16 M^3}
\big\{ i D_0, [ \vDel \cdot \vE] \big\}
\nn \\
&&- 
\frac{c_D}{16 M^3}
\big[ i D_0, \vD \cdot \vE + \vE \cdot \vD \big]
\Bigg]
\phi
\label{eq:intermed}
,\eeq
where we have rewritten the relativistic parameters in terms of the non-relativistic ones.
To arrive at the above form, 
we utilized the identity 
\beq
\big\{ D^i ,  [ \partial_0  E^i ] \big\}
= 
\big[ D_0, \vD \cdot \vE + \vE \cdot \vD \big]
-
2 i Z \vE^2
,\eeq
to remove an operator involving the time derivative of the electric field.

To arrive at the HQET Lagrange density from Eq.~\eqref{eq:intermed}, 
we need to invoke field redefinitions. 
As we have seen,
care must be applied in field redefinitions to preserve the time dependence of Green's functions. 
Defining the 
$\cO(M^{-2})$
equation of motion operator 
$\Box$
by
\beq
\Box
&=& 
i D_0
+ 
\frac{\vD^2}{2M}
+
c_D
\frac{ [ \vDel \cdot \vE]}{8 M^2}
\label{eq:eomop}
,\eeq
we see that a large class of such field redefinitions take the general form
\beq
\phi 
= 
\left[ 
1 
+
\sum_{j=1} f_j  \Box^j \right]
\phi'
\label{eq:general}
,\eeq 
where 
$f_j$
are arbitrary coefficients. 
Because the Green's function 
$G' (x,y)$
of the redefined field
satisfies the equation 
$\Box_y G' (x,y) = i \delta (x - y)$, 
the Green's function of the original field is related by
\beq
G(x,0)
= 
\left[
1
-
i
\sum_{j=1}
f_j \Box^{j-1} \delta(x)
\right]
G'(x,0)
,\eeq
and will not be altered aside from singular behavior at the point 
$x_\mu = 0$.

To produce
Eq.~\eqref{eq:eomop}
as the equation of motion operator for the redefined field 
$\phi'$, 
we require that the field redefinition have the explicit form
\beq
\phi 
= 
\left[
1
-
\frac{\Box}{4M}
+
\frac{3 \Box^2}{32 M^2}
- 
\frac{5 \Box^3}{128 M^3} 
\right]
\phi'
.\eeq
In terms of the redefined field
$\phi'$, 
the Lagrange density to
$\cO(M^{-3})$
becomes
\beq 
\cL 
&=&
\phi'^\dagger 
\Bigg[
i D_0 
+ 
\frac{\vD^2}{2M}
+ 
c_D
\frac{[ \vDel \cdot \vE]}{8 M^2} 
+ 
\frac{\vD^4}{8 M^3}
\nn \\
&&
+
\frac{i c_D}{16 M^3}
\big\{
D^i, 
[ \vDel \times \vB ]^i
\big\}
+
c_{A_1}
\frac{\vB^2 - \vE^2 }{8 M^3}
\nn \\
&&
-
c_{A_2}
\frac{\vE^2}{16 M^3}
- 
\frac{D_0\vD^2 D_0}{4 M^3}
\nn \\
&&
+ 
\frac{- \frac{1}{2} c_D - Z}{8 M^3}
\big[ i D_0, \vD \cdot \vE + \vE \cdot \vD \big]
\Bigg]
\phi'
\label{eq:almost}
,\eeq
which is nearly identical to the HQET Lagrange density in 
Eq.~\eqref{eq:nonrel}.
The last term appearing above is precisely the equation of motion operator we retained in HQET, 
with a coefficient
$c_{X_0}$, 
moreover,
which agrees with that obtained by matching Green's functions in uniform electric fields, 
see Eq.~\eqref{eq:missing}. 
The second-to-last term is a new operator. 
Unlike the last term, 
however, 
this new operator can be removed by a field redefinition without altering the time dependence of Green's functions. 
The requisite field redefinition does not fall into the class considered above in 
Eq.~\eqref{eq:general}. 
In fact,  
we must take
\beq
\phi' 
= 
\left( 
1 
-
\frac{\vD^2}{8M^3} \Box
\right)
\phi''
,\eeq
so that the Lagrange density rewritten in terms of the field 
$\phi''$
is missing the 
$D_0 \vD^2 D_0$
operator. 
This final field redefinition allows us to see the simple relation between Green's functions
\beq
G'(x,0)
= 
\left[
1
+   
\frac{i \vD^2}{8 M^3} \delta(x)
\right]
G''(x,0)
,\eeq
which is valid up to 
$\cO(M^{-4})$.
Aside from a singular contribution at 
$x_\mu = 0$, 
there is no modification to the Green's function. 
Alternatively we can treat the new operator appearing 
Eq.~\eqref{eq:almost}
in perturbation theory to compute the Green's function, 
and arrive at the same conclusion. 
Nonetheless, 
brute-force expansion of the relativistic scalar action confirms the matching conditions determined above.

\section{Summary} \label{s:S}              %

Our primary goal is to understand matching in effective field theories 
with the inclusion of classical external fields. 
Understanding effective field theories in external fields is particularly relevant for lattice QCD computations of certain hadronic observables. 
On the surface, 
there are inconsistencies between effective field theory matching of 
$S$-matrix 
elements,
and the external field correlation functions which should depend on the same physical parameters.

We uncover a potential stumbling block for effective field theories in classical external fields 
in Sec.~\ref{s:demo}. 
In an external field with arbitrary spacetime dependence, 
we find that the effective field theory must include operators related by equations of motion. 
Lacking an on-shell condition, 
the Green's functions are the only theoretical constructs available
to extract physical parameters of the effective theory. 
The Green's functions, 
however, 
are altered by the field redefinitions necessary to remove operators related by equations of motion.
Starting with the most general effective field theory including such operators, 
one cannot pass to the reduced theory. 
As a result, 
the Green's functions generally depend on unphysical parameters which must be isolated to extract physical couplings of the effective field theory.

Despite this general obstruction, 
we consider the particularly simple case of the uniform external field problem for a charged scalar hadron, 
such as the pion. 
This is undertaken in Sec.~\ref{s:scalar}. 
Due to the simplicity of the action in uniform electric fields, 
we demonstrate that operators related by equations of motion do not alter the spacetime dependence of Green's functions.
Because of this fortuitous situation, 
the relativistic effective theory can be written down without including operators related by the equations of motion. 
The same is not true in the non-relativistic effective theory of the charged scalar.

Applying equations of motion to reduce the non-relativistic theory leads to inconsistencies, 
see Eq.~\eqref{eq:striking}. 
The culprit of these inconsistencies is the field redefinition required to remove operators that
are ordinarily redundant. 
Even in uniform electric fields, 
the required field redefinition in the non-relativistic theory alters the Green's function in an essential way, 
see Eq.~\eqref{eq:relationism}. 
Retaining operators related by equations of motion necessitates additional matching conditions. 
The coefficients of equation-of-motion operators cannot be determined from matching 
$S$-matrix elements.
One must appeal to matching at the level of the Green's function, 
which contains information beyond that entering on-shell processes. 
Using the Green's function computed in the relativistic effective field theory provides a way to determine coefficients of operators related by equations of motion in the non-relativistic theory. 
Expanding the relativistic Green's function in an external electric field, 
we obtain the coefficient of an equation-of-motion operator in the non-relativistic theory.  
This Green's function matching is exhibited by employing both HQET and NRQED power counting, 
for which we obtain identical results in 
Secs.~\ref{s:EHQET} and \ref{s:ENRQED}, 
respectively. 
An ultimate check of our results is achieved by performing a brute-force expansion of the relativistic Lagrange density. 
This expansion explicitly allows us to track the field redefinitions necessary to arrive at the non-relativistic theory. 
The brute-force expansion in 
Sec.~\ref{s:NRRQED} 
confirms the necessity of including an equation-of-motion operator, 
as well as verifies the value determined for its coefficient.

In our investigation, 
we additionally determine new results that should be useful to lattice QCD computations in external fields. 
To remove potential surface terms in the uniform field problem, 
we formulate the effective field theory on a Euclidean torus in 
Appendix~\ref{s:B},
where such terms are absent. 
In the process, 
we determine a closed-form expression for the finite-size artifacts affecting Green's functions due to  electroperiodic boundary conditions, 
see
Eq.~\eqref{eq:FV}. 
This result should prove useful in addressing finite-size effects in lattice QCD. 
Computing the Green's function of a charged scalar in uniform electric fields, 
we employ NRQED power counting, 
and arrive at a functional form that we suspect will be highly useful in fitting lattice QCD data. 
The proper-time integration required in determining the relativistic correlator results in cumbersome numerical fits;
however, 
the NRQED expansion gives a systematically improvable result for the correlator that does not require
numerical integration, 
see Eq.~\eqref{eq:nonrel_prop_E}. 
We intend to learn whether the NRQED approach is beneficial. 
Finally 
our study focuses exclusively on the case of a charged scalar for simplicity. 
The phenomenologically relevant case of spin-half hadrons must be treated in a similar fashion.
Having exposed the technical challenges, 
we leave this case to future work.

\appendix

\section{Euclidean Torus} \label{s:B}                             %

In this Appendix, 
we investigate the charged particle propagator on a torus in order to remove potential surface terms that could arise for uniform external fields.  
We note that the case of a charged particle propagator in a uniform magnetic field has already been considered on a torus in%
~\cite{Tiburzi:2012ks}.
Thus we restrict our attention to the case of a uniform electric field in Euclidean space, 
which requires only a simple generalization of the magnetic periodic group, 
see~\cite{AlHashimi:2008hr} for clear exposition of the magnetic case.

To compute the charged particle propagator, 
we work in the sector of vanishing transverse momentum, 
$\vec{p}_\perp = (p_1, p_2) = (0,0)$. 
The $x_3$-direction is taken to have length
$L$, 
while the temporal direction has length
$\beta = 1 /T$.
On a Euclidean torus, 
the four-vector potential
$A^{FV}_\mu = - \mathcal{E} x_4 \delta_{\mu 3}$
is not periodic, 
however, 
it is periodic up to a gauge transformation
\begin{equation}
A^{FV}_3(x + \beta \hat{x}_4) 
=
A^{FV}_3(x)
+
\partial_3 \Lambda (x)
,\end{equation}
where 
$\Lambda(x) = - \mathcal{E} \beta x_3$. 
The gauge transformed scalar field then obeys what we call electroperiodic boundary conditions
\begin{eqnarray}
\Phi^{FV}(x + L \hat{x}_3) 
&=& 
\Phi^{FV}(x),
\notag \\
\Phi^{FV}( x + \beta \hat{x}_4)
&=&
e^{ - i Z \mathcal{E} \beta x_3} 
\Phi^{FV}(x)
.\end{eqnarray}
Consistency of these boundary conditions requires quantization of the field strength,
namely
$Z \mathcal{E} \beta L = 2 \pi n_\phi$, 
where 
$n_\phi$
is the flux quantum of the torus%
~\cite{'tHooft:1979uj,Smit:1986fn,Damgaard:1988hh}. 
The boundary conditions satisfied by the fields ensure that there are no surface terms for any gauge invariant operators appearing in the action.

The finite volume propagator which satisfies the appropriate electroperiodic boundary conditions 
can be constructed from images of the infinite volume propagator. 
Explicitly we have
\begin{eqnarray}
\cG_\cE^{FV}(x',x)
&=&
\frac{1}{L}
\sum_{\nu, n_3}
e^{- 2 \pi i n_3 (x'_3 - x_3)/L}
e^{ i Z \mathcal{E} \beta \nu x'_3}
\notag \\
&&
\times
\cG_\cE
\left(
x'_4 + \nu \beta - \frac{n_3}{n_\phi} \beta, 
x_4 - \frac{n_3}{n_\phi} \beta 
\right) \label{eq:FVAA}
, \quad \end{eqnarray}
where the sum over the temporal winding number
$\nu$, 
and sum over the periodic momentum index 
$n_3$
both extend from 
$- \infty$
to 
$\infty$. 
Above we use the notation 
\begin{equation} \label{eq:AA}
\cG_\cE (\tau', \tau)
=
\frac{1}{2} 
\int_0^\infty ds \,
e^{ - \frac{1}{2} s E^2_\cE }
\langle \tau', s | \tau, 0 \rangle_\cE
,\end{equation}
with
\beq
\langle \tau', s | \tau, 0 \rangle 
&=& 
\sqrt{\frac{Z \cE}{2 \pi \sinh Z \cE}}
\notag \\
&& \times
e^{ - \frac{Z \cE}{2 \sinh Z \cE s} \left[ (\tau'^2 + \tau^2) \cosh Z \cE s -  2 \tau' \tau \right] }
.\eeq
Notice that 
$\cG_\cE (\tau,0)$
determined from 
Eq.~\eqref{eq:AA}
agrees with 
Eq.~\eqref{eq:rel_prop_E}.
To simplify the finite volume propagator in
Eq.~\eqref{eq:FVAA}, 
we set 
$(x_3, x_4) = (0,0)$
and
perform the compact integral over 
$x'_3$, 
with 
$x'_4 = \tau$. 
This procedure yields
\begin{equation}
\cG^{FV}_\cE(\tau)
=
\sum_\nu
\cG_\cE
\left(
\tau, - \nu \beta 
\right) \label{eq:FV}
,\end{equation}
where we define
$\cG^{FV}_\cE(\tau) = \int_0^L dx'_3 \, \cG^{FV}_\cE (x',0)$.
In this form, 
we see that the sector of zero winding number, 
$\nu = 0$,
corresponds to the infinite temporal extent limit employed in the main text above. 
Contributions with non-zero winding number correspond to finite-size corrections, 
and these vanish exponentially with 
$\beta$. 
In practice, 
these corrections are useful to know in order to address finite-size effects on correlation functions calculated with lattice QCD.

In the non-relativistic theory, 
the field 
$\phi^{FV}$
satisfies a variant of electroperiodic boundary conditions by virtue of 
Eq.~\eqref{eq:Phi_vs_phi}, 
namely%
\footnote{
One must be careful to note the conjugate field does not satisfy the complex conjugate of the boundary conditions. 
Instead, we have
$\phi^{FV ^\dagger} (x + L \hat{x}_3) 
= 
\phi^{FV \dagger}(x)$
and
$\phi^{FV \dagger} (x + \beta \hat{x}_4) 
= 
e^{- \beta M} 
e^{i Z \mathcal{E} \beta x_3} \phi^{FV \dagger} (x)$. 
}
\begin{eqnarray}
\phi^{FV}(x + L \hat{x}_3) 
&=& 
\phi^{FV}(x),
\notag \\
\phi^{FV}( x + \beta \hat{x}_4)
&=&
e^{ \beta M}
e^{ - i Z \mathcal{E} \beta x_3} 
\phi^{FV}(x)
.\end{eqnarray}
As a consequence, 
all gauge invariant operators in the NRQED action are periodic, 
and one need not worry about surface terms. 
The non-relativistic correlation function can be constructed from electroperiodic images 
as was done for the relativistic correlator. 
The final result for the non-relativistic 
$x'_3$-integrated correlator
$G_{\cE}^{FV} (\tau) = \int_0^L d x'_3 \, G^{FV}_{\cE} (x',0)$
is given by
\begin{equation}
G^{FV}_{\cE}(\tau)
= 
\sum_\nu
e^{ - \nu \beta M}
G_{\cE}(\tau, - \nu \beta)
,\end{equation}
which is quite similar to Eq.~\eqref{eq:FV} above.

\section{NRQED Expansion of the Relativistic Propagator} \label{s:A}      %

One of the alternate ways to determine the parameter 
$c_{X_0}$,
which enters the non-relativistic effective theory, 
is to start with the fully relativistic scalar propagator in an electric field and perform the NRQED expansion. 
To perform the NRQED expansion of Eq.~\eqref{eq:rel_prop_E}, 
we note the velocity scaling of the parameters
\begin{eqnarray}
\mu &=& E_\cE \, \tau \sim \cO(v^{-2}), 
\notag \\
\zeta &=& Z \cE \tau^2 \sim \cO(v^{-1})
.\end{eqnarray}
We must be careful about sub-leading corrections, 
however, 
because the initial energy
$E_\cE = M + \frac{1}{2} 4 \pi \alpha_E \cE^2 + \cdots$
contains a subleading term which contributes at 
$\cO(v^{4})$
to the parameter
$\mu$.
Using the relation between the relativistic and non-relativistic scalar fields in 
Eq.~\eqref{eq:Phi_vs_phi}, 
the relativistic 
$\cG_\cE$
and non-relativistic 
$G_\cE$
Euclidean two-point correlation functions are in turn related by
\beq
G_\cE(\tau)
=
2M
e^{M\tau}
\left[
1
+
\frac{\zeta^2}{\mu^2}
\right]^{1/4}
\cG_\cE(\tau).
\label{eq:G_vs_cG}
\eeq
To aid in the velocity expansion, 
we rescale the proper time to be dimensionless, 
$\cs=Z\cE s$. 
After this rescaling, 
we have the relativistic propagator in the form
\beq
\cG_\cE(\tau)
=
\frac{1}{2\sqrt{2\pi Z\cE}}\int_0^\infty d\cs
~\textrm{exp}
\left[
-\frac{1}{2}f(\cs)
\right],
\eeq
where 
\begin{equation}
f(\cs)
=
\frac{\mu^2}{\zeta}
\cs
+
\zeta \coth \cs
+
\ln \sinh \cs
.\end{equation} 
Because terms in the exponent become large and negative as 
$v\rightarrow0$, 
the bulk of the integrand arises from 
$\cs$ 
where the exponent is a maximum. 
This maximum occurs at the value
\beq
\coth \cs_0
=
\frac{
1
+
\sqrt{1+4 ( \zeta^2 + \mu^2)}
}
{2\zeta}
> 0
.\eeq

Using Laplace's method, 
we expand 
$f(\cs)$ 
about 
$\cs_0$,
and 
the relativistic correlator can be written in the form
\beq
\cG_\cE(\tau)
&=&
\frac{e^{- \frac{1}{2} f(\cs_0)}}
{\sqrt{2\pi Z \cE \, f''(\cs_0)}}
\int_{-\frac{ \cs_0}{2}  \sqrt{f''(\cs_0)}}^\infty
dS~e^{-S^2}
\nn 
\\
&&
\times 
\exp
\left[
-\frac{1}{2}
\sum_{j=3}^\infty 
\frac{f^{(j)}(\cs_0)}{j!}
\left(\frac{ 2 S}{\sqrt{f''(\cs_0)}} \right)^j
\right].
\nn \\
\eeq
The lower bound of integration has the form
$-\frac{\cs_0}{2}\sqrt{f''(\cs_0)} = - \sqrt{\frac{\mu}{2}} + \cdots \sim \cO(v^{-1})$, 
and can therefore be extended to 
$-\infty$ 
up to corrections that are exponentially small. 
To compute the propagator to 
$\cO(v^4)$
accuracy, 
we require 
$j_\text{max} = 6$. 
Performing the Gaussian integration, 
and then expanding to this order, 
we find
\begin{eqnarray}
\cG_\cE(\tau)
&=& 
\frac{e^{- \left(\mu + \frac{\zeta^2}{6 \mu} \right)}}{2 M}
\Bigg[
1
- 
\frac{\zeta^2}{4 \mu^2}
\left( 1 - \frac{\zeta^2}{10 \mu} \right)
\notag \\
&&- 
\frac{\zeta^2}{4 \mu^3}
\left( 
1 
-
\frac{5 \zeta^2}{8\mu} 
+ 
\frac{17 \zeta^4}{280 \mu^2}
-
\frac{\zeta^6}{800 \mu^3}
\right)
+\cO(v^6)
\Bigg]
.
\nn\\
\end{eqnarray}
Finally appending the conversion factors in 
Eq.~\eqref{eq:G_vs_cG}
and expanding to the same order, 
we obtain the non-relativistic correlator
\begin{equation}
G_E(\tau)
=
e^{
- 
\frac{\zeta^2}{6 \mu} 
+ 
\frac{\zeta^4}{40 \mu^3}
-
\frac{\zeta^6}{112 \mu^5}
-
\frac{\zeta^2}{4 \mu^3}
- 
(E_\cE - M) \tau
}
,\end{equation} 
which is exactly the same as 
Eq.~\eqref{eq:nonrel_prop_E}
with the replacement
$c_\text{NR} \to c_\text{R}$.

\begin{acknowledgments}
This work is supported in part by the Alfred P.~Sloan foundation through a CUNY-JFRASE award, 
and by the U.S.~National Science Foundation, under Grant No.~PHY$12$-$05778$.
The work of BCT is additionally supported by a joint CCNY--RBRC fellowship. 
We thank 
A.~Walker-Loud
for discussion, 
and 
R.~Hill for correspondence that incited this investigation. 
\end{acknowledgments}

\appendix

\bibliography{matching}

\end{document}